# Detecting Home Locations from CDR Data: Introducing Spatial Uncertainty to the State-of-the-Art

**Maarten Vanhoof**
Open Lab, Newcastle University,
Newcastle-upon-Tyne, UK.
Orange Labs,
Paris, France.

M.vanhoof1@newcastle.ac.uk
Maarten.vanhoof@orange.com

**Fernando Reis**
Eurostat
Luxembourg, Lux

**Zbigniew Smoreda**
Orange Labs,
Paris, France

**Thomas Plötz**
Open Lab, Newcastle University,
Newcastle-upon-Tyne, UK

## ABSTRACT
Non-continuous location traces inferred from Call Detail Records (CDR) at population scale are increasingly becoming available for research and show great potential for automated detection of meaningful places. Yet, a majority of Home Detection Algorithms (HDAs) suffer from "blind" deployment of criteria to define homes and from limited possibilities for validation. In this paper, we investigate the performance and capabilities of five popular criteria for home detection based on a very large mobile phone dataset from France (~18 million users, 6 months). Furthermore, we construct a data-driven framework to assess the spatial uncertainty related to the application of HDAs. Our findings appropriate spatial uncertainty in HDA and, in extension, for detection of meaningful places. We show how spatial uncertainties on the individuals' level can be assessed in absence of ground truth annotation, how they relate to traditional, high-level validation practices and how they can be used to improve results for, e.g., nation-wide population estimation.

### Author Keywords
CDR data; Meaningful Place Detection; Home Detection, Spatial Uncertainty

## INTRODUCTION
Location aware computing represents one of the pillars of ubiquitous computing [43] and ever since Weiser formulated the vision of the third generation of computing [44] substantial progress has been made with regards to both sensing physical or geographical location as well as to utilizing location information for context aware applications [6,17].

With the widespread availability of mobile sensing platforms, such as the prevalent smartphone that is equipped with GPS or GSM based location sensing, outdoor localization can be considered solved [46]. The availability of high resolution, outdoor localization capabilities has enabled very sophisticated location aware applications (e.g., support of visually impaired people in public transport access [33]) but also allows personalized analysis and analysis of movement patterns [21].

Yet, the detection of meaningful places – this is, locations where persons spend a significant amount of their time with most prominent examples being home or work [29] – based on sensed location traces often requires active participation of users. If not for annotation of locations then the users are solicited for the activation of tracking or at least through their consent for sharing exact location data, which for many means a burden or raises privacy concerns [12].

As an alternative, research has engaged with passive surveying of travel to subsequently derive meaningful places. GPS based applications, for example, allow for extraction of meaningful places from trajectory traces[1] by, for instance, detecting trip purpose (e.g., home-work commuting). Works on small scale samples have achieved promising results [5,13,20,28,45] but their limited scope of observations – both in space, time, and number of users, still limits location sensing for many applications and analysis techniques [12]. For example, planners or epidemiologists would not only require small samples of location traces but rather large-scale recordings of movement patterns, ideally at population scale.

With large samples in mind and in addition to the analysis of GPS sensed traces, efforts have been made to exploit essential system information routinely recorded by mobile phone network operators for billing or maintenance purposes (e.g., [9,30]). A prominent example of such low-level information are *Call Detail Records (CDR)*, that is attributes of phone calls that include connection time, duration, source and destination number, and respective cell-tower IDs for both caller and callee, providing location information on both. CDR data are very attractive for large-scale analysis of location and movement patterns even though they are typically very sparse given that they only contain information from the time of a call. The attraction of the data lies in the fact that they are routinely recorded, network-wide and as such provide direct access to localization data for large samples of populations

---

[1] For an excellent review on the use of GPS technology for travel surveys see [35]).

Similar to GPS-based location traces, CDR data have been analyzed with the objective of identifying meaningful places in a person's life. Through analyzing meaningful places at population-wide scale conclusions can be drawn with regards to, for example, planning, policy making, or infrastructure investments [2]. Beyond such pragmatic applications, in-depth sociological studies have been conducted based on the extraction of meaningful places from CDR data (e.g., [41], [31])

In this paper, we review meaningful places detection based on CDR data thereby focusing on the most important case of automated home allocation. We argue that beyond the works performed on small-scale continuous traces, the majority of works on CDR data face several methodological challenges. Most notably, we find that Home Detection Algorithms (HDA) for large-scale data lack ground-truth data at the individual level to develop learning methods or evaluation criteria. As a consequence, the majority of HDA use simple and implicit criteria for the semantic annotation of user traces on which no consensus nor assessment of sensitivity exists in literature. Additionally, we observe that assessments of errors for HDA are restricting themselves to nation-wide comparison with census data, since no framework has been developed that target lower levels in absence of ground truth. All of this fundamentally limits the utility of current HDA and limits the potential of large dataset for systematic exploitation.

Based on an exemplary, very large-scale dataset of CDR recorded in France, we rigorously investigate the effects of five different criteria used in current HDAs. We compare findings between the different criteria and use a unique validation dataset with a much finer resolution than before to compare results to groundtruth. In a second step we construct a refined, unified framework to create a data-driven assessment of spatial uncertainty for home detection on individual traces. We calculate the different measures for spatial uncertainty for all of the five home detection and investigate their temporal and spatial properties. We found high correlations between our spatial uncertainty measure and validation results for all algorithms. We show how incorporating spatial uncertainty even in simple algorithms drastically improves results.

Our findings present the first work in appropriating spatial uncertainty in HDA and, in extension, for detection of meaningful places based on CDR data analysis. We show how spatial uncertainties on the individual level can be assessed in absence of ground truth annotation, how they relate with high-level validation practices, and how they can be used to improve results for nation-wide population estimation. With this framework, large-scale explorations of location and movement patterns become more robust and validation practices can be complemented.

## RELATED WORK

In context-aware computing location is by far the most researched source of contextual information [29]. However, raw location data is often of limited use for research, policy makers or the development of Location Based Services (LBS) if no meaning for individual users can be added. Thus, detection of meaningful places from location data becomes important. Meaningful places hereby refers to locations where a person spends a significant amount of their time or where he/she performs particularly relevant activities, the most prominent examples being home or work place [1,29].

The importance of detecting meaningful places, in particular home, for large-scale populations is prevalent in a lot of different research areas and applications. Since meaningful places form focal points of spent time, social activity, mobility, economic consumption and labour (re)generation, their detection for large-scale populations will render important insights in the wider structures governing our society. One application area is official statistics where the place of residence is a key concept in population statistics used for establishing voting rights and allocating budget [42]. The place of residence also serves as a starting point for establishing migration, population mobility and tourism statistics [16].

### Change in tracking technology: GPS to CDR

With the advent of ubiquitous computing technologies, research efforts have focused on accessing a wide range of digital traces to render detection of meaningful places completely passive and automated, preferably for large populations. A naïve distinction can be made between works that use continuous and non-continuous traces even though the transition between both is difficult to fix.

Works on analyzing continuous user traces were at the forefront of early developments and used small-scale samples most commonly from continuous GPS traces but also from Bluetooth, or wireless LAN positioning. For example, [45] explored possibilities to derive trip purposes from GPS traces based on GIS-analysis as early as 2001; finding amplification of their idea in the slow replacement of travel diaries by GPS surveys in transport research throughout the decade [35]. The general methodology consists of a two-step approach in which clustering of location traces to detect important places leads to results mostly by means of time-space heuristics [28]. Techniques to detect places from continuous traces range from GIS analysis [20,45] to popular k-means algorithms [4], non-parametric Bayesian approaches [28], and even fingerprinting of the radio environment [23] and can be applied to a range of digital traces, sometimes even including non-continuous traces [28].

The increasing relevance of non-continuous traces became prominent through growing efforts in harvesting new datasources consisting of large amounts of location traces from, amongst others, mobile phone usage, credit card transactions and check-ins on location based services (e.g., Foursquare or Gowalla) or online social networks (e.g., Twitter). Despite consisting much sparser traces, enthusiasm for those new datasources has been rising because of much larger coverage in terms of users, timespan and spatial extent.

**Detecting meaningful places from CDR data**

In this paper we focus on CDR data as one example of such newly available, non-continuous location traces focusing on home detection as an example of detecting meaningful places. According to the wide consensus in the literature CDR data can be considered the most relevant of these datasources. However, our analysis is – strictly speaking— not limited to this data source as they are generalizable in principle to the entire group of large-scale, non-continuous location traces.

CDR data capture the phone activity of subscribers on the operator's network. They consist of records that store information about location (the cell-tower used), temporality (time and duration of usage), and interaction (who contacts whom). CDR data are by definition non-continuous as information is only stored when a call or text is made or received. Penetration rates of mobile phones are estimated to a staggering 96% worldwide (2014; [24]), indicating the potential to sense billions of people both in developed as emerging economies. In addition, mobile phone usage has become widely accepted and possible in almost every situation or environment. Compared to most other data gathering location traces, CDR data is considerably more cost-effective, less biased (due to usually large sample sizes and widespread adaptation), less limited in scale, and less limited in data resolution. [26,27].

Besides applications on nation-wide communication networks [22,37], CDR datasets have widely been mined for its mobility traces to study, amongst others[2], large scale human mobility patterns (e.g., [27,32,36]) and population presence (e.g., [15,26]). Over the years, several empirical findings indicate the strong potential for detection of meaningful places from CDR data. Most notably, power-law-like distributions of human displacement, mobility motifs, visitation frequencies and staying times all relate to the idea that human mobility, when studied based on CDR data, is predictable to a very high degree and across all classes of population as people tend to spent most of their time in few locations [14,21,34,38,39].

Having unveiled the potential of CDR data to capture important locations, i.e., locations where persons spend most of their time, substantial research focuses on detection of meaningful places. [1] was one of the first to investigate the possibilities of deriving meaningful locations for individual users from CDR data. Note that detecting homes based on CDR data actually refers to detecting the cell-tower that operates in the area of the home location. For a CDR dataset in Estonia, a multi-stepwise model has been proposed for the detection of home, work and multifunctional anchor points. Based on experiments with the individual traces of 14 users, the detection algorithms uses spatial (grouping of adjacent cell-towers) and temporal (average starting times of call, and standard deviations of starting times of calls) criteria. Validation of the 282,572 detected home anchor points was done by comparison with the Estonian Population Register. Validation of the work-time and multifunctional anchor points has not been done due to the lack of validation data at the investigated scale.

In [25] home and work locations were identified from CDR data in urban areas in the USA (Los Angeles and New York City). Individual traces from CDR data have been clustered by means of the Hartigan algorithm to detect important locations [10]. For validation scores from a logistic regression model have been used. That model was constructed on a dataset of 18 volunteers and incorporates criteria like active days, distinct days, duration of calls, events during working hours and during 'home' hours.

The limitations of these works lie in their dependence on very small samples for training. To overcome this problem, [18] used a training-set of 5,000 users to learn the CDR-based behavioral fingerprint for residential locations in an emerging economy. Home locations of the users in the set are derived from their permanent contract with the provider, which in their case is only available for 5% of the entire CDR dataset. In developed countries, the possibility to create such training-sets does not exists due to the legal obligation on anonymizing users. For this reason, [14] used an unsupervised k-means algorithm with features derived from mobile phone user behavior to cluster the frequent locations for 100,000 users in Portugal. Annotation of the obtained clusters, mostly based on average temporal patterns, allows for the detection of home, work and unidentified locations on a large-scale base.

So far only little progress has been made to successfully tackle the small training-set problem. In fact, the general tendency is to use rather simplified home allocation algorithms that incorporate generic criteria for home detection instead of using criteria derived from training-sets. Such methodology has widely been applied in works that use home location for a large sample of users as a prerequisite for further analysis [8]. For CDR data, such applications exist that study human mobility, epidemiology, social interactions, or economic development, all of which dependent on the detection of home (e.g., [19,27,31,40]).

A first critique of the state-of-the-art is that existing works use a definition of home implemented already during the detection of important places. Annotation thus implicitly happens *a priori* by means of the choice for home criteria, rather than by being derived from training-sets. The most popular criteria for home definitions are temporal limitations during nighttime ('home is the location with most activity between x pm and y am'), but also temporal aggregations (distinct days, or even weekend-days) and spatial grouping (most observations within a spatial reach) are being deployed (e.g., [27,31,40]). As an example, [11] uses the highest distinct number of observations between 6pm and 8am as criteria to derive home locations from a Boston dataset, basing the chosen time-interval on statistics from the American Time Use Survey.

---
[2]For an extensive literature review on the different applications of CDR data in research, see [7].

Although these criteria seem rather logical, it is remarkable that no research exists on the sensitivity of results – neither on nation-wide level, nor over time, nor in space and also not for different datasets, with the only exception being [8] where the effects of using different criteria on a credit card transactions and a Flickr database in Spain have been evaluated. As a consequence there is no consensus on which criteria to use best.

To support the use of their, almost arbitrary, criteria, research has embraced validation by census data. Given the population wide coverage of CDR datasets this is not surprising but the choice to use census data as validation is often taken for granted and implications are rather poorly discussed. Using census data as validation dataset for large-scale home detection –be it by comparing population densities or through derived commuting figures— is, strictly speaking, a rather limited alternative solely justified by the absence of real validation data, which typically does not exists at that scale. In-fact, census data has never specifically been gathered to serve as validation dataset for home detection. Consequently, it is difficult to assess how 'valid' census data actually is and to which degree it (mis)shapes methodologies and understanding.

In addition, census data only allows for high-level validation of the developed methodologies. Validating HDAs on a nation-wide level provides little to no information on the accuracy at individual or meta-level. As a consequence, "[n]ot a single study has systematically validated the methods used to derive activity locations from those passively generated mobile phone datasets" [12]. This shortcoming limits the choice of methods that rely on learning. It also results in unknown errors and uncertainty in studies that use meaningful places as a prerequisite.

In the next sections we elaborate on these critiques, using a large scale CDR dataset for France, which we consider representative for other large-scale non-continuous datasets. We construct and apply five different HDAs, all incorporating one specific criterion to define home. We compare findings between the different criteria and use a unique validation dataset created in close cooperation with the French National Statistics Office (INSEE) to perform a conventional third party validation – however at a much larger spatial resolution. In a second step we construct a framework to assess spatial uncertainty for home detection on individual traces. We compute the different measures for spatial uncertainty at individuals' level for all of the five HDAs. While investigating the temporal and spatial properties of our newly constructed spatial uncertainty measures, we show how they can complement traditional high-level validation by census. In a next step we explore the potential of using spatial uncertainties for performing data-driven assessment of HDA performance, so eliminating the necessity for third party validation. In a final step we experiment with the incorporation of spatial uncertainty information in simple HDAs, showing their potential to improve results on a nation-wide scale.

## DATA

Call Detail Record (CDR) data are collected by mobile phone service providers for billing and network maintenance purposes. Being collected every time a call or text is initiated, they store locational (the used cell-tower) temporal (time and duration of usage) and interactional (who contacts whom) information for both correspondents. Location traces from CDR data thus are non-continuous as they are user initiated and rather sparse in time. Previous research on CDR data has been made possible by providers in Western-Europe and the United States mostly (Belgium, France, Portugal, LA, Boston, etc.), but also African (Côte d'Ivoire, Senegal) and Asian (China) datasets are becoming available. In compliance with ethical and privacy guidelines CDR data are anonymized and, for recent datasets, often aggregated or re-anonymized every given time-period.

### French CDR data

In this paper, we use access to an anonymized CDR dataset recorded by Orange[TM]. The data covers mobile phone usage of ~18 million subscribers on the Orange network in France during during a period of 154 consecutive days in 2007 (May 13, 2007 to October 14, 2007). Mobile phone penetration being estimated at 86% [3] at that time and given a population of 63,945 inhabitants during the observed period[3], that is a rough 32% of all French mobile phone carriers and 28% of the total population.

The Orange[TM] France 2007 CDR dataset is one of the largest CDR datasets available worldwide in terms of population-wide coverage and has been extensively studied (e.g., [15,22,37]). It is the latest CDR dataset available for France that allows for such a long term, continuous temporal —but anonymised— tracking of users in France. More recent datasets are limited by The French Data Protection Agency (CNIL) who are anticipating the EU General Data Protection Regulation and do not allow to collect individual traces as they are considered risky even if personal identification information is irreversibly recoded.

The spatial accuracy of the dataset is restricted to the network's spatial resolution, i.e., to the locations of the cell towers as installed by the network provider. The spatial distribution of the 18,273 cell-tower locations is known but not uniform. In general, higher densities of antennas are found in more densely populated areas like cities or coastlines. Lower densities of cell towers are observed in more rural areas, as well as in mountain or natural reserve areas. The Voronoi tessellation of all cell tower locations is shown in figure 4 illustrating the coverage and the density of the network.

---
[3] This is the average of the monthly estimates for the period between Mai and October 2007 as obtained from the French National Statistics Website (www.insee.fr)

The temporal resolution of the analysed dataset is inhomogeneous as CDR are only created and stored during calls thereby generating records on both caller and callee side. For example: for one arbitrary day of the covered timespan (Thursday, $1^{st}$ October 2007), the median number of records per user was four, relating to only two different locations. Such statistics are representative for CDR based studies and can be deemed rather high compared to other large-scale non-continuous datasets like credit-card transactions or Flickr photos [8]. In total, 65% of our location traces is related to call activity, while the remaining 35% is related to text.

Temporal sparsity in observations —only a few records per user per day— and spatially inhomogeneous distributions of covered areas —as a result of demand-driven, non-uniform distribution of antenna locations— are typical characteristics of CDR datasets and pose substantial challenges for their automated analysis. On the contrary, the very large scale reach at population level without requiring active participation of the user for location sharing whilst at the same time preserving anonymity as well as privacy is very attractive for many application areas. Aggregating data over extended periods of time enables complex analysis and diminishes influence from singular events and/or non-routine behaviour.

**National Statistics Validation dataset**
The CDR dataset is anonymized and does not contain any personalized or detailed information other than the elementary, technical information as outlined before. In order to validate HDAs on a nation-wide scale a population density dataset was created in close collaboration with the French National Statistics (INSEE). To construct this validation dataset, the Public Finances Directorate General (DGFIP) collected individual (or household) locations from revenue declarations, the housing tax and the directory of taxable individuals. Subsequently, the French National Statistics office calculated population densities at the Orange network level by aggregating home locations to the nearest cell-tower.

Unfortunately, such large-scale, high resolution home location information could only be made available to us for the year 2010. We opt to use this validation dataset over the low resolution, publicly available census data as the spatial translation of statistical sectors to cell-tower coverage. Given the spatially inhomogeneous distribution of cell-tower locations this translation is complicated and prone to errors (see, e.g., [18]). With this in mind having access to a validation dataset at the same spatial resolution as the mobile phone network is a huge advantage as it has both a higher resolution of information and a higher accuracy. Note that in our analysis we only use the validation dataset for relative comparisons of the analyzed home allocation methods, i.e., no absolute validation is attempted. We, however, do make the assumption that, at a nation-wide level, relative population patterns do not change drastically within three years.

**HOME ALLOCATION ANALYSIS FOR CDR DATA**

Based on a literature review that covered, for example (but not exclusively), [1,11,12,14,25,27,31,40], we identified five different criteria that are often used when detecting homes from CDR data. To compare how results change with different criteria, as well as to explore the spatial uncertainty measures for different criteria, we construct five algorithms, each incorporating a specific criterion. The five different algorithms define home as the location where:

1. The majority of both outgoing and incoming calls and texts was made;
2. The maximal number of distinct days with phone activities —both outgoing and incoming calls and texts— was observed;
3. Most phone activities were recorded during 7pm and 9am;
4. Most phone activities were recorded, implementing a spatial perimeter of 1000 meter around a cell-tower that aggregates all activities within and
5. The combination of 3) and 4) thus most phone activities recorded during 7pm and 9am and implementing a spatial perimeter of 1000 meter.

Note that some of these algorithms are identical to works where home locations were used as a prerequisite. These criteria, however, were also used in more elaborated works (e.g., [14,25]). In addition, [8] uses a similar methodology and similar criteria when comparing home detection based on credit card transactions and Flickr data. The relevance of these algorithms thus goes further than CDR data alone.

We apply all five aforementioned algorithms to the Orange$^{TM}$ France 2007 CDR dataset, aiming for detecting presumed home allocations (L1) for all users during all months in the dataset (May-October). Besides the actual annotated home we gather several metadata on the workings of the algorithms on each case. Most importantly we also gather information about the second (L2) and the third (L3) most plausible locations for home as defined by the different algorithms. In the remainder of this paper we will refer to these locations as L2, and L3, with L1 being the actual detected home location by a particular algorithm.

**Comparisons of results at individual level**

To compare results between two different home detection methods at the individual level, we assess to which degree a set of algorithms detects the same location as home. A straightforward approach to do so is through evaluating the Simple Matching Coefficient – SMC [8]:

$$SMC\ (algorithm_A, algorithm_B) = \frac{\sum_{i=1}^{N} \delta(Home_{A,i}, Home_{B,i})}{N}$$

where $i=1..N$ denotes the $N$ users analysed, and $\delta(Home_{A,i}, Home_{B,i})$ is the Kronecker delta that equals to *1* when the home detected by algorithm A for the *i*-th user is identical to the home detected by algorithm B for the same user. It is *0* otherwise. Values of SMC range between *0* and *1* and can easily be interpreted as the percentage of individual cases for which both algorithms detected the same home locations.

**Validation of results at cell-tower level**

To compare results from the home allocation methods with the proposed validation dataset we evaluate the degree of similarity in population numbers attributed to the different cell-towers on a nation-wide scale. Note that we do not target an absolute assessment of similarity, as this is impossible given the unknown spatial distribution of the 28% sample of Orange users and the differences in times of collection between the CDR dataset and our validation data. Instead, we compare general patterns of estimated populations by means of vector comparison. In our case, a first vector x will denote the estimated population by an algorithm for all cell-tower areas and is compared to a second vector y that describes the validation population for exactly the same cell-tower areas. Both vectors thus have an equal length (representing the 18,273 cell-towers in the Orange network). To quantify (dis-)similarities of aforementioned two vectors, we use a standard Cosine Similarity Metric (CSM):

$$\cos(\vec{x}, \vec{y}) = \frac{\vec{x} \cdot \vec{y}}{||\vec{x}||\ ||\vec{y}||}$$

Values of the cosine will range between -1 and 1. A value of 1 indicates the highest similarity in orientation (the angle between x and y is zero degrees), 0 indicates the lowest similarity in orientation (the angle between vector x and vector y is 90 or -90 degrees) and -1 indicates an opposite orientation (the angle between x and y is 180 degrees)

Deriving the angle between two vectors and expressing it in degrees (°) thus becomes:

$$CSM\ (\vec{x}, \vec{y}) = |\cos^{-1}\left(\frac{\vec{x} \cdot \vec{y}}{||\vec{x}||\ ||\vec{y}||}\right) * \frac{180}{\pi}|$$

A CSM value of 0° denotes the highest possible similarity, 90° indicates the lowest similarity in orientation whereas 180° degrees refers to an opposite orientation.

**Spatial uncertainty**

Probably the most important step in detecting home is the decision to annotate one of the detected important places as home. Since most of the time no ground truth data is available at the individual level (neither for learning, nor for validation), the error of erroneous decisions is unknown. Here, we propose a way to calculate the uncertainty that comes with deciding for one home location, given that other important places derived from the same individual traces (L2, L3) could also, plausibly, be the home location.

We construct our measure of uncertainty to be explicitly spatial − Spatial Uncertainties. The idea behind is that if distances between the three top locations (L1, L2, and L3) are small, the spatial error when annotating the wrong location as home will remain small. In this case, the spatial uncertainty of home detection can be considered small. If distances are high, however, an erroneous detection of home will result in a high spatial error. Home detection in that case corresponds to a high spatial uncertainty.

To construct our measure of spatial uncertainty, we compare the plausibilities of different locations to be home by calculating their ratio of amount of observations that have passed the criteria by the deployed algorithm. If the criteria of an algorithm, for example, demand the highest amount of distinct days for detecting home, then the comparison of plausibility will be on the amount of distinct days at the different considered locations. We explicitly incorporate the spatial extent of the uncertainty by inserting the absolute distances between considered locations. Distances between locations (e.g., uncertainty because of long distance travelling) and differences in spatial resolutions of observations (e.g., high-density cell-tower areas versus low-density) will therefore both resonate in the proposed spatial uncertainty measure:

$$SU_n = \sum_{\{i,j,\ldots\}} \frac{p_i}{p_n} * \frac{d(n,i)}{2}$$

where $SU_n$ is the spatial uncertainty for detecting a home in location n (in meters), i and j are the first denoters in a set of other possible locations for home, $p_n$ and $p_i$ are the number of considered observations, given the criteria in the algorithm used, in respectively location n and i, and d(n,i) is the distance between locations n and i (in meters).

The spatial uncertainty of a location *n* is influenced by the distance to all other considered locations and by the share of evaluated observations in these locations compared to the amount of evaluated observations in location *n*. A smaller share of evaluated observations in other locations and smaller distances to these other observations both result in smaller spatial uncertainties, indicating a higher plausibility that home detection is done at the correct location.

We limit the calculation of SU to detected home locations only, i.e., $n = 1$. Note that, in principle, the formula also allows to calculate the spatial uncertainty related to deciding on a second, third, etc. plausible location. However, without loss of precision (see below) we limit our analysis to only two other plausible locations, limiting the set of $\{i, j, ...\}$ to i and j only. Doing so, the SU measure that we will use can be written as:

$$SU_{L1} = \frac{p_{L2}}{p_{L1}} * \frac{d(L1,L2)}{2} + \frac{p_{L3}}{p_{L1}} * \frac{d(L1,L3)}{2}$$

Applying this formula to a simplified example where we use algorithm 1 (which considers the total amount of activities) to detect the home of user x who has a trace of calling 10 times at location A and 5 times at location B and 1 times at location C. The distances between location A, B and C are all 1 km. Based on the criteria used by algorithm, location A is L1, B is L2 and C is L3. The according SU for detection home at location A then becomes:

$$SU_{L1} = \frac{5}{10} * \frac{1000}{2} + \frac{1}{10} * \frac{1000}{2} = 300 \text{ meter}$$

## RESULTS

We have applied five HDAs, all incorporating different criteria, to the French CDR dataset. We analyzed their reciprocal differences and performances based on our validation dataset. In a next step we discuss the construction of spatial uncertainty measures for the five algorithms. Building on this, we show the potential of using spatial uncertainty measures for a data-driven evaluation of HDAs in absence of validation datasets. Ultimately, we show how incorporation of spatial uncertainty by means of a very basic filtering mechanism leads to better results.

### Numbers of Users analyzed

We applied all five HDAs to each of the six months available in the French dataset. For each algorithm we detected homes for ~18 million users per month, resulting in a maximum of ~109 million cases for each algorithm. Remember that with detecting home, we mean, detecting the cell-tower that covers the area of the home location. Table 1 shows the exact numbers and the amount of cases for which a second or third plausible locations was evaluated.

|  | *Algo 1* | *Algo 2* | *Algo 3* | *Algo 4* | *Algo 5* |
|---|---|---|---|---|---|
| *Number of detected homes* | 109.4 (100%) | 109.4 (100%) | 98.4 (100%) | 109.4 (100%) | 98.4 (100%) |
| *Cases with L2* | 102.2 (93.5%) | 102.2 (93.5%) | 78.0 (81.3%) | 102.0 (92.8%) | 78.4 (79.6%) |
| *Cases with L3* | 96.1 (87.9%) | 96.1 (87.9%) | 65.0 (66.1%) | 94.7 (86.6%) | 62.3 63.3)% |

**Table 1: Total numbers of detected homes by different algorithms and amount of cases that had plausible L2/L3 locations. Percentages are given in brackets.**

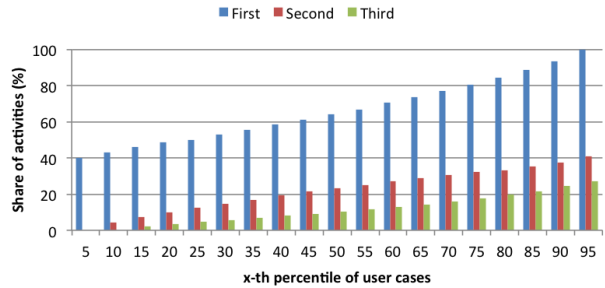

**Figure 1: Percentile distributions of shares of activities in top 3 most frequent locations for all 110 million detection cases.**

Depicting 66%, to 86% of cases with (at least) three plausible locations, these numbers show how loose the used criteria for home detection are and thus why the assessment of uncertainty deemed necessary. Given competitive locations for home detection, we explored the importance of locations in terms of observed activity. For each user, we calculated the share of activities performed in the top 3 most frequently used cell-tower. The percentile distribution of the shares for the whole French dataset is given in figure 1. The importance of the location with most observations is clearly visible. Its shares range from ~40% (5[th] percentile) to 100% (95[th] percentile) with a median of 64%. Shares of the second and third most frequently visited cell-towers are substantially lower with medians of 23% and 10% and a 95[th] percentile of 41% and 27%. These figures support our decision to limit the analysis of spatial uncertainty to only three plausible locations. Observations are similar to [14] that found 95% of the users in a Portugal CDR dataset have fewer than four frequent locations.

### Comparing detected homes for different algorithms

To tackle the question to which degree different algorithms detect equal home locations, we calculated pairwise SMC values. When calculating SMC values, we omit all cases where one of the considered algorithms failed to detect a home location (e.g., when no observations where left after implementing a time constraint). Figure 2 shows the SMC values for all algorithm combinations and for different months. Degrees of accordance range between 61.5% and 96.4% of the detected homes, resulting in discordance rates between 40% and 4%, which translate in absolute numbers of 6.8 and 0.6 million users.

The patterns of (dis)similarities between algorithms are clearly visible. Algorithms that incorporate time-constraints (algorithms 3 and 5) disaccord to high degrees with algorithms that count amount of activities (algorithm 1), distinct days (algorithm 2), or perform spatial grouping (algorithm 3), all of which show high degrees of pairwise accordance. The criteria of time-constraints thus results in different detected homes for 30% to 40% of the cases compared to all other criteria. Possibly these differences sterns from sparser observations but also different spatial behavior during nighttime could form an explanation.

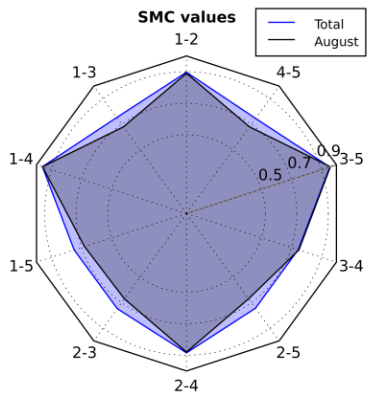

**Figure 2:** Radar plot showing comparisons of the number of identical homes detected for August and the total dataset by means of pairwise SMC values for the different combinations of algorithms 1 – 5. Other months display similar values.

**Validation with census data: CSM**

In order to validate the performance of the analyzed algorithms, we compare numbers of detected home locations aggregated at cell-tower level to corresponding numbers of our validation dataset by means of the Cosine Similarity Metric (CSM). Figure 3 shows the calculated CSM values for all algorithms and for different months.

The distinct days criteria performs best in replicating the population pattern of the validation dataset, followed by the number of activities and the time-restraining number of activities. The criteria that involve grouping in space perform worst, even though the applied criterion (1 kilometre) is not that far.

Temporal patterns are similar for all algorithms, with lower CSM values for June and September, and higher values for May, July, August and October. A possible explanation for high SMC values for May and October is the limited number of available data days for these months in the dataset (18 and 14 days), which supports our choice to analyse the dataset on a monthly basis.

Highest SMC values are observed during summer (July and August). All algorithms are sensitive to this temporal change, probably, because of changing spatial behaviour of users that go on holidays [15]. Time-restraining criteria are more sensitive for these changes which questions their widespread adaptation in literature. In addition it is an interesting observation that differences between algorithms are smaller than differences of similar algorithms trough time. Future deliberation on algorithm choice, we suggest, should therefore take into account time-effects or thus, more in general, the characteristics of the deployed dataset.

CSM values between 35° and 38° are still far off the intended 0° as perfect matching with the validation set would achieve. In light of this, differences between algorithms or months that comprise 1° or 2° degrees do not seem relevant. However, when looking at the spatial patterns related to such small changes in CSM values, their relevance becomes clear.

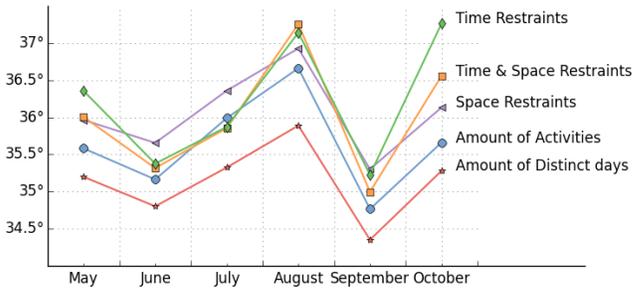

**Figure 3:** CSM values (in degrees) for the comparison between amount of detected homes per algorithm and the validation ground truth data, both at the cell-tower level.

Figure 4, for instance, shows the spatial patterns, emphasised in "hot" (marked red) and "cold" (marked blue) spots, of the detected homes in June and August by the amount of activities algorithm. The difference in CSM values between June and August is 1.08° but results in a very different spatial pattern, pronouncing the holiday trips of French people that direct to coasts and mountainous areas.

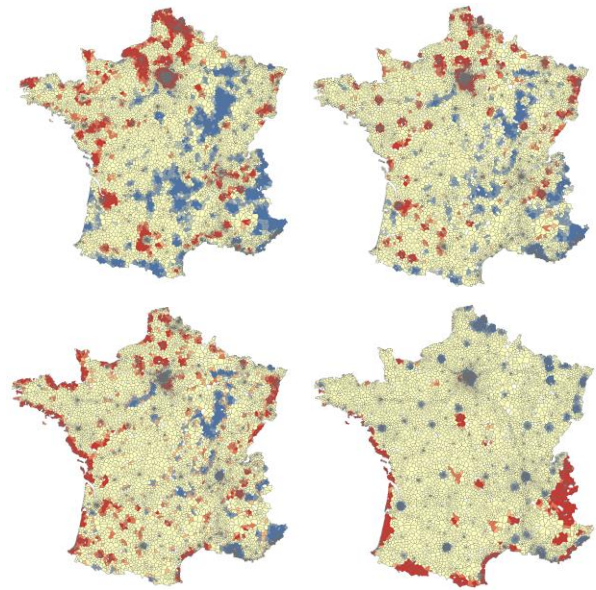

**Figure 4:** Hotspots (red) and coldspots (blue) defined by the 90+% interval of the Getis-Ord Gi* statistic for population numbers of the validation dataset (a), the number of detected homes by the amount of activities algorithm in June (b) and August (c) and the median SU values for detected homes by the amount of activities in August (d). The map is a made up by the Voronoi tessalation based on cell-tower locations

**Spatial uncertainty construction**

In order to evaluate the spatial uncertainty related to the home detection choice of the different algorithms, we calculated the SU values for home location of each treated case. The medians of the SU figures aggregated for all five algorithms and all six months are shown in figure 5. The observed SU values range between 2.5 km and 7 km, suggesting that generally, spatial uncertainty on the detected homes is moderate.

The number of activities criterion depicts the lowest SU values. Remarkably, the SU of maximum activities and time-constraining algorithms are almost equal during non-summer months. However, during summer, SUs of the time-constraint criteria increases drastically, whereas the SUs of the maximum activities only rise moderately. Again, this poses questions w.r.t. the use of time-constraints algorithms, especially during summer months.

Despite having the lowest CSM values, the distinct days criterion has the highest SU values during non-summer months. This is an understandable result as limiting the maximum observations in a month to the number of distinct days (normally 30 or 31) will give more weight to the L2 and L3 locations. A clear indicator that using distinct days can be risky as it might favor secondary locations.

The higher SU values in summer suggest a change in the nature of observed traces. Home detection is more uncertain due to higher distances between plausible locations and different calling patterns at these locations; a change poorly dealt with in existing algorithms given their growing discordance with the validation dataset during July and August (see figure 4 (d) where the spatial pattern of calculated SU-values for the amount of activities criteria in August can be investigated). In addition, low SU values in May and October suggest no change in the nature of traces and so the observed high CSM values for these months can not be explained in a similar way.

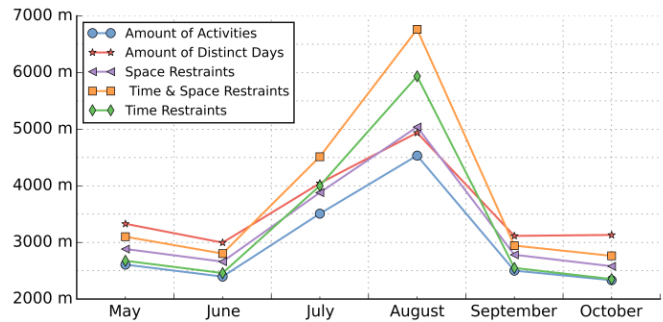

**Figure 5: Medians of calculated Spatial Uncertainties for detected homes by five algorithms during different months.**

The similarity in temporal patterns between SU and CSM values suggests that spatial uncertainties at the individual level are linked to the nation-wide performance of the algorithms. Figure 6 illustrates this relation and shows a strong correlation between both measures (R=0.46 to R=0.68 depending on the omittance of outliers for May and October). This is an important finding as it opens the door for a data-driven assessment of different home algorithms, so diminishing, or even excluding the role of external validation datasets.

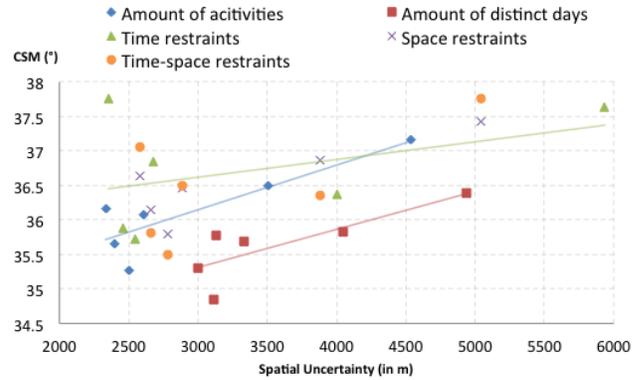

**Figure 6: Relation between SU and CSM values (median per month per algorithm). Trendlines are given for the amount of activities, amount of distinct days and time-restraint criteria. A positive, general correlation with R= 0.46 is obtained for all observations when omitting observations for time-restraints and time-space restraints algorithm in October (outliers in the left top corner). A positive correlation with R=0.68 is found when omitting all observations for October and May.**

Since SU and CSM values are correlated, we were interested in whether we could use information on SU at the individual level, to improve performance of algorithms on the nation-wide scale. One simple way would be by using SU values as a filter, thus discarding individual traces that depict too high spatial uncertainties. We therefore investigate the sensitivity of CSM values to filtering on different parameters of SU.

Figure 7 shows the resulting CSM values for the amount of activities and time-restraint algorithms when filtering on SU-values lower than 10, 30, 50 or 70 km. The results exceed our expectations. Filtering on SU values significantly improves the performance of HDAs regardless the threshold set for SU. In addition, SU filtering seems to eliminate, to a large degree, the effect of summer months in our dataset since obtained CSM values became more constant in time. On the downside, SU filtering seems to have limited effects for the months May and October.

**CONCLUSION**

The analysis of CDR data has great potential for automated detection of meaningful places. However, the validation of locations for individual users remains a challenge given the absence of ground truth for such large-scale datasets. As a consequence, previously developed methods based on continuous location traces not necessarily scale for large scale datasets. Yet, research has continued to use home detection from CDR data, mostly as a prerequisite for further research, but largely relying on rather simple Home Detection Algorithms based on implicit criteria for the definition of home. Despite widespread application, no consensus on the use of these criteria exists and neither does a critical assessment of their application to CDR data.

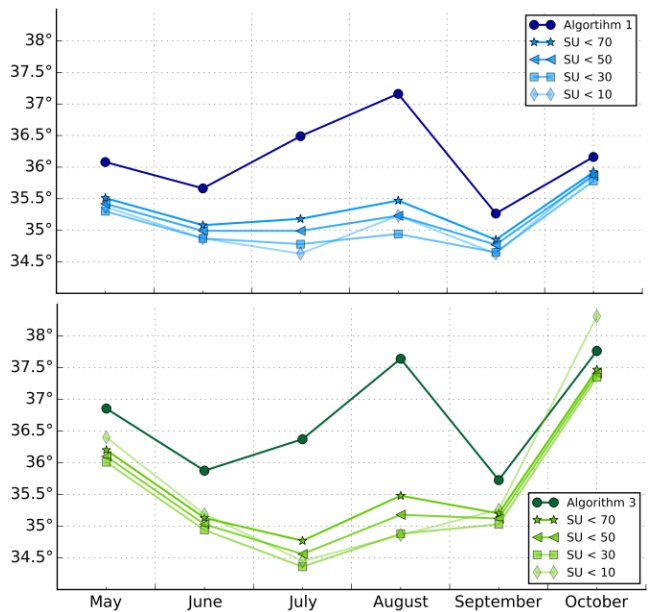

Figure 7: CSM values (in degrees) for the amount of activities criteria (top) and the time-restraint criteria (bottom) and their application when adapting filtering on SU-values (expressed in km) lower than 70, 50, 30 and 10 km. Effects of SU-filtering for the other algorithms show similar results.

Comparing findings from the application of five simple HDAs, each based on a popular criterion to a French CDR dataset, showed large differences between criteria (up to 40% of the considered cases), large differences in performance, and a huge sensitivity to time of observations, especially during summer months. Third party validation of the large-scale patterns of population density in France based on a unique dataset prepared by the French Statistical Office showed the 'distinct days'-criterion to perform best.

Given the absence of proper ground truth annotation, validation of HDAs for CDR dataset is mostly based on a very high level comparison with census data. To waive such high-level aggregated estimation of error, we proposed the construction of a measure of spatial uncertainty (SU) for home detection on individual location traces. Calculation of SU-values was discussed and applied for five algorithms on the French CDR dataset. Interestingly, the aggregation of SU-values at cell-tower level correlated strongly with previous performance measures obtained from third party validation (between R=0.46 and R=0.68) and generalizes for all assessed algorithms. This opens possibilities to develop SU for data-driven assessment of HDAs, to complement traditional validation methods to become more multi-level and adaptive in time, and to limit the dependency of methods to existing validation datasets.

We experimented with the use of SU to improve population estimation based on HDA. For each of the five algorithms, we deployed a SU-filter, so using the calculated risks on spatial error associated with individual location traces. For each of the algorithms, results outperformed the simple HDAs deployed before regardless the parameters used for filtering. Surprisingly, SU filtering seems to eliminate, to a large degree, the effect of summer months in the performance of our algorithms. We believe these findings to be relevant for further research in the detection of meaningful places from CDR data. Most notable, we believe that the calculation and comparison of SU for different data-algorithm combinations, the multi-level and spatio-temporal analysis of SU, the integration of SU in the learning phase of more complex HDAs and the development of a data-driven assessment of performance based on SUs are promising directions of future work.